# Geometric approach of degree of polarization in 3D polarimetry

Yahong Li,[1,2] Yuegang Fu, [1,2,*] Wenjun He,[1,2] Zhiying Liu, [1,2] Jianhong Zhou,[1] Yu Zhao, [1,2] P. J. Bryanston-Cross,[3] and Yan Li[4]

[1]School of Opto-Electronic Engineering, Changchun University of Science and Technology, Changchun, China
[2]Key Laboratory of Opto-Electric Measurement and Optical Information Transmission Technology of Ministry of Education, Changchun University of Science and Technology, Changchun, China
[3]School of Engineering, University of Warwick, Coventry CV4 7AL, UK
[4]Department of science and technology, Changchun University of Science and Technology, Changchun, China
*fuyg@cust.edu.cn

The geometric approach presented in this paper concerns the degree of polarization ($DoP$) of a random 3D statistical electromagnetic field. We use that a 3×3 coherency matrix can always be decomposed into an incoherent superposition of three specific coherency matrices, to construct the first orthant of a spherical shell ( $0 < x, y, z < 1, \sqrt{3}/3 \leq r \leq 1$ ). For a random 3D statistical electromagnetic field, the polarimetric result of the $DoP$ can be expressed in terms of a point located in the first orthant of this spherical shell physically. Furthermore, based on the intrinsic relationship between the defined parameters ( $DoP_{3D-p}$, $DoP_{3D-pp}$ ) and $DoP$, a spatially quadric surface is obtained to quantify the values of the $DoP$ for all physically reachable points contained in the first orthant of this spherical shell. Within this geometric approach, two examples are cited to demonstrate the applicability of 3D polarimetry to intuitively display the polarimetric result of a random 3D statistical electromagnetic field.

**Keywords**: Polarization, Statistical electromagnetic fields, Ellipsometry and polarimetry

Recently, the three-dimensional (3D) polarization is of topical interest in 3D statistical electromagnetic fields [1-6]. In the 3D polarization cases, the usual 2D mathmatical and geometric representations of polarization are not available [7-10]. The necessity of introducing appropriate approaches to represent the polarization properties of 3D statistical electromagnetic fields. One of the important physical quantities is the degree of polarization ( $DoP$ ) for 3D statistical electromagnetic fields. The concept of $DoP$ was firstly introduced by Samson [11], and it was developed by Barakat [12], T. Setälä [13] and José J. Gil [14] et al. Now, the definition of DoP has been accepted that two parameters are needed for specifying the $DoP$ of a random 3D statistical electromagnetic field: indices of polarimetric purity ($P_1$, $P_2$) named by José J. Gil [15-17].

Based on the definition of $DoP$, Colin J. R. Sheppard and José J. Gil have proposed geometric representations of the $DoP$ in terms of a triangular composition plot [10] and a polarimetric purity space [15,16]. However, the main aim of this paper is based on the above-mentioned parameters ($P_1$, $P_2$) to introduce a completely different geometric approach, in which the $DoP$ of a random 3D statistical electromagnetic field can be represented by a fixed point in the first orthant of a spherical shell ( $0 < x, y, z < 1, \sqrt{3}/3 \leq r \leq 1$ ). At the same time, using the intrinsic relationship between the $DoP$ and the defined parameters ( $DoP_{3D-p}$, $DoP_{3D-pp}$ ), a spatially quadric surface is depicted at the below of the first orthant of this spherical shell. Because the two spatial graphics geometrically form a strict downwards projection, it is appropriate for 3D polarimetry to quantify the DoP of a 3D statistical electromagnetic field.

In 3D polarimetry [18,19], the polarization properties of a random 3D electromagnetic field can be expressed in terms of nine measurable real values,

$$\mathbf{S}_{9\times 1} = (s_0, s_1, s_2, s_3, s_4, s_5, s_6, s_7, s_8)^T, \qquad (1)$$

where the first parameter $s_0$ is the total intensity of the measured 3D electromagnetic field, the others $s_j$ ( $j = 1,...,8$ ) are the intensity values of the measured 3D electromagnetic field along the characteristic directions determined by eight 3×3 Gell-Mann matrices [20].

For a random measured 3D electromagnetic field, accordong to the inherent transformation relationship between the Eq. (1) and the 3×3 coherency matrix [7], the measured 3D electromagnetic field can be calculated,

$$\Phi = \begin{pmatrix} \frac{1}{2}s_3 + \frac{\sqrt{6}}{6}s_0 + \frac{\sqrt{3}}{6}s_8 & \frac{1}{2}(s_1 - i \cdot s_2) & \frac{1}{2}(s_4 - i \cdot s_5) \\ \frac{1}{2}(s_1 + i \cdot s_2) & \frac{\sqrt{6}}{6}s_0 + \frac{\sqrt{3}}{6}s_8 - \frac{1}{2}s_3 & \frac{1}{2}(s_6 - i \cdot s_7) \\ \frac{1}{2}(s_4 + i \cdot s_5) & \frac{1}{2}(s_6 + i \cdot s_7) & \frac{\sqrt{6}}{6}s_0 - \frac{\sqrt{3}}{3}s_8 \end{pmatrix},$$

(2)

where $\Phi$ is the 3×3 coherency matrix, and it is a positive semidefinite Hermitian matrix. The calculated 3×3 coherency matrix contains all polarization properties of the measured 3D electromagnetic field.

In order to better analyze the polarization properties of the measured 3D electromagnetic field, we decompose the 3×3 coherency matrix shown in Eq. (2) into a incoherent superposition of three specified polarized components, the DoP values of which are 1, 0.5 and 0, respectively. The decomposition result of the 3×3 coherency matrix can be expressed,

$$\Phi = I_{3D\text{-}p} \cdot \Phi_{3D\text{-}p} + I_{3D\text{-}pp} \cdot \Phi_{3D\text{-}pp} + I_{3D\text{-}up} \cdot \Phi_{3D\text{-}up}, \quad (3)$$

and

$$\begin{cases} \Phi = (\nu_1, \nu_2, \nu_3) \cdot Diag(\lambda_{01}, \lambda_{02}, \lambda_{03}) \cdot (\nu_1, \nu_2, \nu_3)^\dagger, \\ \lambda_{0j} = \frac{\lambda_j}{\lambda_1 + \lambda_2 + \lambda_3} \ (j=1,2,3, \lambda_1 \geq \lambda_2 \geq \lambda_3 \geq 0), \\ I_{3D\text{-}P} = \lambda_{01} - \lambda_{02}, \Phi_{3D\text{-}P} = \nu_1 \otimes \nu_1^\dagger, \\ I_{3D\text{-}pp} = \lambda_{02} - \lambda_{03}, \Phi_{3D\text{-}pp} = (\nu_1 + \nu_2) \otimes (\nu_1 + \nu_2)^\dagger, \\ I_{3D\text{-}up} = \lambda_{03}, \Phi_{3D\text{-}up} = (\nu_1 + \nu_2 + \nu_3) \otimes (\nu_1 + \nu_2 + \nu_3)^\dagger, \end{cases}$$

(4)

where $\lambda_j$ and $\nu_j$ ($j=1,2,3$) are the eigenvalues and eigenvector of the 3×3 coherency matrix shown in Eq. (2). The subscripts $_{3D\text{-}p}$, $_{3D\text{-}pp}$, and $_{3D\text{-}up}$ represent the 3D totally polarized component, 3D partially polarized component and 3D totally unpolarized component, respectively. $I_{3D\text{-}p}$, $I_{3D\text{-}pp}$ and $I_{3D\text{-}up}$ are the normalized intensity values, $\Phi_{3D\text{-}p}$, $\Phi_{3D\text{-}pp}$ and $\Phi_{3D\text{-}up}$ are the 3×3 coherency matrices of three specified polarized components. The symbol $\otimes$ indicates the Kronecker product, and $\dagger$ represents the complex conjugate transpose.

The above decomposition method included in Eqs. (3)-(4) had also been mentioned in several papers [15-17], and it was named as the characteristic or trivial decomposition. In some literatures [1,15], the second term $\Phi_{3D\text{-}pp}$ with $DoP$ of 0.5 was defined as a two-dimensional (2D) unpolarized conponent, it is not always ture. A detail discussion of this explanation will be included in the last two examples.

To quantify the weight of the specified three polarized components in the measured 3D electromagnetic field, we define three parameter,

$$\begin{cases} DoP_{3D\text{-}p} = \frac{I_{3D\text{-}p}}{I_{Total}} = \frac{\lambda_1 - \lambda_2}{\lambda_1 + \lambda_2 + \lambda_3}, \\ DoP_{3D\text{-}pp} = \frac{I_{3D\text{-}pp}}{I_{Total}} = \frac{2(\lambda_2 - \lambda_3)}{\lambda_1 + \lambda_2 + \lambda_3}, \\ DoP_{3D\text{-}up} = \frac{I_{3D\text{-}up}}{I_{Total}} = \frac{3\lambda_3}{\lambda_1 + \lambda_2 + \lambda_3}. \end{cases}$$

(5)

Obviously, the value ranges of three parameters are between 0 and 1, and the three parameters are always satisfied with the identical equation,

$$DoP_{3D\text{-}p} + DoP_{3D\text{-}pp} + DoP_{3D\text{-}up} = 1. \quad (6)$$

In the literatures [15,16], José J. Gil had defined $P_1 = DoP_{3D\text{-}p}$ as the degree of purity, and $P_2 = 1 - DoP_{3D\text{-}up}$ was defined as the degree of directionality.

Based on the inequality of arithmetic and geometric means, the three parameters meanwhile fullfill the following two inequations,

$$\begin{cases} DoP_{3D\text{-}p}^2 + DoP_{3D\text{-}pp}^2 + DoP_{3D\text{-}up}^2 = r^2, \\ r^2 \geq \frac{(DoP_{3D\text{-}p} + DoP_{3D\text{-}pp} + DoP_{3D\text{-}up})^2}{3}, \\ r^2 \leq (DoP_{3D\text{-}p} + DoP_{3D\text{-}pp} + DoP_{3D\text{-}up})^2. \end{cases}$$

(7)

According to Eqs. (6) and (7), the final expression can be rewritten as,

$$1/3 \leq r^2 \leq 1. \quad (8)$$

In order to better demonstrate the applicability of the above proposed geometric approach, we combine with the $DoP$ for 3D electromagnetic field defined in the literatures [3,4], the relationship among the defined three parameters shown in Eq. (5) and the DoP is derived,

$$DoP^2 = DoP_{3D\text{-}p}^2 + \frac{1}{2} \cdot DoP_{3D\text{-}p} \cdot DoP_{3D\text{-}pp} + \frac{1}{4} \cdot DoP_{3D\text{-}pp}^2. \quad (9)$$

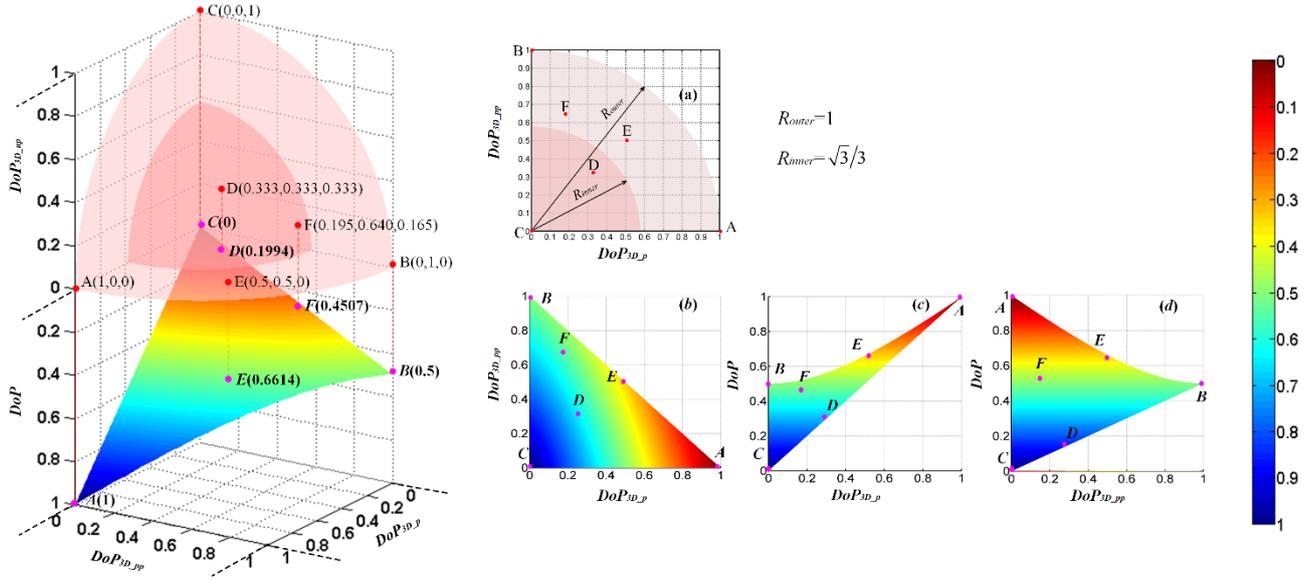

FIG. 1 The firs orthant of a spherical shell ($0 < x, y, z < 1, \sqrt{3}/3 \leq r \leq 1$) is constructed to show the polarimetric result of the $DoP$, in which the x-axis, y-axis and z-axis are defined as the parameters: $DoP_{3D\text{-}p}$, $DoP_{3D\text{-}pp}$ and $DoP_{3D\text{-}up}$. In order to quantify the value of the $DoP$ for the polarimetric result, a spatially quadric surface is depicted at the below of the first orthant of this spherical shell, in which the x-axis, y-axis and z-axis are specified as the parameters: $DoP_{3D\text{-}p}$, $DoP_{3D\text{-}pp}$ and $DoP$. Besides, the subgraph (a) is the top view of the first orthant of this spherical shell, and the subgraphs (b)–(d) are the orthogonal 2D projections of the spatially quadric surface on the: (b): x-y plane, (c): x-z plane and (d): y-z plane, respectively.

Consequently, using the above defined three parameters $DoP_{3D\text{-}p}$, $DoP_{3D\text{-}pp}$ and $DoP_{3D\text{-}up}$ in Eq. (5) as the coordinate axes of a space coordinate system $\{x, y, z\}$, the first orthant of a spherical shell ($0 < x, y, z < 1, \sqrt{3}/3 \leq r \leq 1$) is constructed, as shown in Fig. 1. Similarly, using the relationship among the parameters $DoP_{3D\text{-}p}$, $DoP_{3D\text{-}pp}$ and $DoP$ included in Eq. (9), a spatially quadric surface is obtained at the below of the first orthant of this spherical shell. Here, it is worth noting that the x-axis ($DoP_{3D\text{-}p}$) and y-axis ($DoP_{3D\text{-}pp}$) of the two spatial graphics are completely coincident, and the z-axes ($DoP_{3D\text{-}up}$ and $DoP$) are exactly the opposite. Therefore, the two spatial graphics geometrically form a strict projection relation, as shown in Fig. 1.

Next, we mainly discuss the physically reachable area of the first orthant of this spherical shell. When the first inequation in Eq. (7) takes the equal sign, the values of the three parameters must be satified with the condition that $DoP_{3D\text{-}p} = DoP_{3D\text{-}pp} = DoP_{3D\text{-}up} = 0.3333$. The special case is completely possible for 3D electromagnetic fields to occur during 3D polarimetry. Combined with Eq. (8), it is verified that this special case corresponds to a 3D partially polarized field with a $DoP$ of 0.1944. Hence, it is concluded that only a point on the inner spherical surface ($r = \sqrt{3}/3$) is feasible, as shown as point D (0.3333, 0.3333, 0.3333) in Fig. 1. The color of the position of point D denotes the value of $DoP$.

Similarly, the equal sign of the second inequation in Eq. (7) holds if and only if one of $DoP_{3D\text{-}p}$, $DoP_{3D\text{-}pp}$ and $DoP_{3D\text{-}up}$ is equal to 1. According to Eq. (8), the three cases respectively characterize the totally polarized field, partially polarized field and un-polarized field in 3D. The values of $DoP$ correspondingly are 1, 0.5 and 0. Therefore, the polarimetric results of these three cases are represented by three different points, i.e., the points A (1,0,0), B (0,1,0) and C (0,0,1) in Fig. 1. The color of the position of three points, i.e., blue, light green and red indicate that the values of $DoP$ are 1, 0.5 and 0. Hence, there are three feasible points on the outer spherical surface ($r = 1$).

In addition to the inner and outer spherical surfaces, when the polarimetric point is located in the middle area between the two spherical surfaces ($\sqrt{3}/3 < r < 1$), the measured 3D electromagnetic fields are all partially polarized, and the values of *DoP* can be obtained by making a downwards projection to intersect with the spacially quadratic surface, i.e., the distribution of *DoP* values for all points located in the first orthant of this spherical shell during 3D polarimetry.

Especially, it is worth emphasizing that all polarimetric points will fall in the first orthant of this spherical shell, but all points included in the first orthant of this spherical shell may not be of physical meanings. Obviously, only the point which satisfies the sum of three coordinate values is equal to 1 is physically reachable.

Finally, we take two general 3D electromagnetic fields as examples to demonstrate the applicability and validity of the proposed geometric approach shown in Fig. 1. In our previous work [7], when a high numerical aperture (NA) microscope objective (NA=1.25 immersed in oil) with anti-reflecting (AR) coatings is interacted with a 3D partially polarized field, the sampled exiting fields are expressed in terms of 3×3 coherency matrices,

$$\begin{cases} \boldsymbol{\Phi}(E) = \begin{pmatrix} 0.5 & -0.25i & 0 \\ 0.25i & 0.5 & 0 \\ 0 & 0 & 0 \end{pmatrix}, \\ \boldsymbol{\Phi}(F) = \begin{pmatrix} 0.25 & 0.125 & 0.125i \\ 0.125 & 0.5 & -0.125i \\ -0.125i & 0.125i & 0.25 \end{pmatrix}. \end{cases} \quad (10)$$

Firstly, we make the characteristic decompositions of the 3×3 coherency matrices included in Eq. (10). Then, applied the decomposition results to Eq. (5), the polarimetric points of the exampled 3D electromagnetic fields are determined by E (0.5,0.5,0) and F (0.1952,0.6404,0.1644), which are located in this spherical shell. Finally, we project the points E and F downwards, and we can directly obtain the values of *DoP*, i.e., E (0.6614) and F (0.4507). Therefore, the exampled 3D electromagnetic fields are partially polarized in 3D, and the colors of position E and position F depend on the values of *DoP*, i. e., green and orange, respectively.

Last but not least, we examine the second items $\boldsymbol{\Phi}_{\text{3D-pp}}(E)$ and $\boldsymbol{\Phi}_{\text{3D-pp}}(F)$ in the characteristic decomposition results of these two examples included in Eq. (10),

$$\begin{cases} \boldsymbol{\Phi}_{\text{3D-pp}}(E) = \begin{pmatrix} 0.5 & 0 & 0 \\ 0 & 0.5 & 0 \\ 0 & 0 & 0 \end{pmatrix}, \\ \boldsymbol{\Phi}_{\text{3D-pp}}(F) = \begin{pmatrix} 0.2840 & 0.1213 & 0.216i \\ 0.1213 & 0.4319 & -0.1213i \\ -0.216i & 0.1213i & 0.284 \end{pmatrix}. \end{cases} \quad (11)$$

Obviously, $\boldsymbol{\Phi}_{\text{3D-pp}}(E)$ is a 2D totally unpolarized field, but $\boldsymbol{\Phi}_{\text{3D-pp}}(F)$ is a 3D partially polarized field. Combined with Eq. (10), the first exampled 3D electromagnetic field does not contain z-component, i.e., the vibration directions at a point are statistically in a constant 2D plane at different times. However, the vibration directions of the second example fluctuate in a 3D space, so the second polarized component is no longer 2D unpolarized.

In summary, we have presented a geometric approach to 3D polarimetry that enable illustrate the *DoP* by a point located in the physically reachable area of the first orthant of a spherical shell ( $0 < x, y, z < 1, \sqrt{3}/3 \le r \le 1$ ). Using the downwards projection relation, a quadric surface at the below of the first orthant of this spherical shell is obtained to show each polarimetric point. Because the two spatial graphics geometrically form a strict projection relation, it is very useful for 3D polarimetry to intuitively display the *DoP* of the measured 3D statistical electromagnetic field, and it is also expected to be of application in near-field optics, singular optics and nanophotonics.

We thank the acknowledges support from the National Natural Science Foundation of China (NSFC) (11474037,11474041).